\documentclass[12pt]{article}

\title{What is really ``quantum" in quantum theory?}

\author{Andrei Khrennikov \\International Center for Mathematical
Modeling \\ in Physics and Cognitive Sciences,\\
University of V\"axj\"o, S-35195, Sweden\\
Email:Andrei.Khrennikov@msi.vxu.se}
\begin{document}
\maketitle
\begin{abstract}By analysing probabilistic foundations of quantum theory
we understood that the so called quantum calculus of probabilities (including 
Born's rule) is not the main distinguishing feature of ``quantum". This calculus
is just a special variant of a contextual probabilistic calculus. In particular, we analysed
the EPR-Bohm-Bell approach by using  contextual probabilistic models (e.g., the frequency
von Mises model). It is demonstrated that the EPR-Bohm-Bell consideration are not so much about
``quantum", but they are merely about contextual. Our conjecture is that the ``fundamental quantum 
element" is the Schr\"odinger evolution describing the very special dependence of probabilities
on contexts. The main quantum mystery is neither the probability calculus in a Hilbert space
nor the nonncommutative (Heisenberg) representation of physical observables, but the Schr\"odinger evolution 
of contextual probabilities.
\end{abstract}

\section{Introduction}

Last years there was demonstrated increasing interest to foundations of
quantum theory.\footnote{Intensive development of quantum information theory 
and quantum computing as well as new experimental technologies are the main
stimulating factors.} I would like to mention a few recent investigations
on the general probabilistic structure of quantum theory, L. Ballentine [1], L. Hardy [2],
R. Gill, G. Weihs, A. Zeilinger, M. Zukowski [3], S. Gudder [4], A. Khrennikov [5],  
I. Pitowsky [6], J. Summhammer [7] and on 
the EPR-Bell experiment, L. Accardi [8], W. De Baere [9], W. De Myunck [10],
K. Hess and W. Philipp [11], A. Khrennikov [12], I. Volovich [13]. 

Roughly speaking all these investigations are devoted to the problem formulated in the title
of this paper:

{\bf ``Which element of quantum theory is really the fundamental ``quantum element"?}

The Hilbert space representation of states? Born's rule for probability and, in particular, interference
of probabilities? Noncommutativity of observables?
Heisenberg uncertainty relation? Bohr's complementarity? Reduction?
Schr\"odinger evolution? Or we could not at all splitt quantum theory into 
essential and nonessential elements? At the moment the latter point of view dominates in the quantum
community: quantum theory as indivisible whole.\footnote{However, see S. Gudder [4] and C. Fuchs [14].}

I think that similar problems stimulated recent investigations of G. `t Hooft [15] who
proposed a (discrete) classical deterministic model which beneath quantum mechanics.
I remark that intereference-like statistical effect for macroscopic particles (driven 
by classical electromagnetic forces) was numerically simulated  by A. Khrennikov and
Ja. Volovich [16] in the model with discrete time. 

Mentioned investigations (besides `t Hooft's model) are devoted to the probabilistic structure
of quantum theory. This is very natural, since quantum theory is a {\bf statistical theory.}
This theory can not say anything about individual quantum systems (of course, it depends
on an interpretation  of quantum mechanics). 

{\it Creation of a mathematical theory that would describe dynamics of individual quantum systems
is the greatest problem for physics of this century or even millennium.}

In the present paper I would like to analyse the structure of quantum theory by using 
the so called contextual probabilistic model, see [5]. On the one hand, by such an analysis
we can better understand the probabilistic structure of quantum theory. On the other hand,
by comparing contextual and quantum models we can try to find the fundamental element
of quantum theory.

\section{On exotic theories of probability}

I would like to start with a rather provocative paper [17] of R. Gill, namely, with comments on
the chapter {\it Khrennikov and exotic probabilities.} I totally disagree with R. Gill
in his neglecting of the fundamental role which so called 
``exotic probabilistic models", in particular, CONTEXTUAL probabilistic 
models (and, in particular, von Mises frequency model [18]) can play for clarifying
the probabilistic structure of quantum theory.

{\bf 2.1. ``Classical" and ``quantum" probabilities.} 
If we consider quantum and classical physics from the purely probabilistic
viewpoint then we should recognize that there exist two very different probabilistic
calculi which are used in completely different situations (moreover, they are developed
practically independently):

\medskip

1) {\it Kolmogorov measure-theoretical model,} 1933, [19].

2). {\it Probabilistic calculus in a Hilbert space,} end of 20th (Born, Jordan, Dirac), e.g., 
[20].

\medskip

{\bf Remark.} {\small It looks very  strange (at least for me) that the mathematical 
formalism (Kolmogorov axiomatics)
which describes rigorously classical statistical physics was proposed 
later than the corresponding quantum formalism...}

Since 60th, the main stream of quantum probabilistic investigations was directed
to Bell's model  [21] of the EPR-Bohm experiment [22].\footnote{As pointed out 
I. Volovich [13] we should sharply distinguish  the original EPR model [23] and 
the EPR-Bohm-Bell model [21]. Recently we proved [24] that for the original EPR model
it is possible to construct the local realist representation even by using Kolmogorov-Bell
viewpoint to realism, see our further considerations. I also mention investigations
of S. Molotkov [25] and I. Volovich [13], [26] who demonstrated the fundamental role of
space-time in quantum information theory.} I think that Bell's considerations gave us 
a new reformulation  of the old problem in probabilistic foundations
which was  already observed in the two slit experiment.
In this experiment the conventional rule for the addition of probabilities is perturbed
by the interference term. Already farthers-creators of quantum theory paid large attention
to this difference in probabilistic calculi, see, e.g. Dirac [19] or Feynman [27].

We should do something to solve this problem -- existence of two different probability calculi --
and to unify these models -- Kolmogorov and quantum -- in some way. I
think that such a unification should be done on the basis
of some new intuitively clear probabilistic model. We can not start
directly with a Hilbert space. Of course, such a start would be the simplest from the
mathematical viewpoint. However, it induced (and continues to induce)
terrible misunderstandings, mysteries and prejudices. 

I should recall that there already were a few attempts of such a unification, e.g., models
of S. Gudder [28] and I. Pitowsky [29]. And these models were well done from the mathematical viewpoint!
So formally the problem was solved. The main problem is that those approaches are
not intuitively attractive: e.g., nonmeasurable sets in Pitowsky's model
or so called influence function in Gudder's model. Well, the origin of the {\it influence function} is
little bit less mysterious than the origin of the Hilbert space. But just little bit...

The same problem was studied in huge number of works in quantum logic, see, e.g., [30].
Quantum logic is also well done from the purely mathematical viewpoint. However, intuitively
it is not more attractive than the quantum Hilbert space formalism.

{\bf 2.2. Frequency approach.} I like the frequency model of R. von Mises, [18], because
this is the most intuitively attractive probabilistic approach. I would like to remark that 
the von Mises approach is not at all so bad as it was claimed in 30th, see, e.g., [31], [32],
but here I would not like go into mathematical and logical details.

For me one of the main distinguishing features of the von Mises approach is its 

\medskip
             
\centerline{{\bf              CONTEXTUALITY}}

\medskip

In this approach a complex of physical conditions (in my papers  I propose the terminology - 
physical context or simply {\it context}) is represented by a {\it collective.} R. von Mises underlined:
{\it first a collective then probability.}

In the von Mises model contextualism has the following consequences for the EPR-Bell framework,
see [31]:

\medskip

1) No counterfactual statistical data! 

All statistical data are related to concrete
collectives (contexts).  Here all considerations based on the use of counterfactual arguments (in particular,
counterfactual derivations of Bell's inequality, see, e.g., [33], [34]) are nonsense.\footnote{Similar
anti-counterfactual conclusions were obtained in some other approaches, e.g., W. De Baere [9],
W. De Myunck [10], and K. Hess and W. Philipp [11].}

\medskip

2) New viewpoint of independence.

In the frequency model we use a new notion of independence. Not {\it independence of events}
(as in the Kolmogorov model and some other models), but {\it independence of collectives-contexts.}
R. von Mises strongly criticized the conventional notion of independence, namely, event
independence. He presented numerous examples in which conventional independence 
was represented as just a meaningless game with numbers -- to obtain factorization
of probability into the product of probabilities. In the frequency theory we study
independence of collectives (in my terminology -- contexts). 

If we analyse the well known  Bell-Clauser-Horne locality condition by using independence 
of collectives then we see immediately, see [31],  that corresponding collectives
are dependent because they contain the same particle-preparation procedure. 
Hence there are no reasons to  suppose the validity of this condition or 
to connect this condition with locality.

\medskip

3) Existence of probability distributions for hidden variables.

In the von Mises approach we should start with analysing an experimental situation
to be sure that we really have the statistical stabilization (existence of limits)
of relative frequencies. In the opposite to Kolmogorov's model we can not start 
directly with a probability distribution (as Bell did in his Kolmogorov-version
of the EPR-arguments). And here a rather strange, but important question arise:

{\it Why do we suppose that a Kolmogorov-probability distribution of hidden variables
exists at all?}

Well, our experimental equipment produces the statistical stabilization of relative
frequencies for observables. Why  should it produce the statistical stabilization of
relative frequencies for hidden variables? I do not see such reasons... Moreover, intuitively
it looks that we can not provide such a statistical stabilization for microsystems: they
are too sensible to our macroscopic preparation procedures. In principle, relative frequencies
can fluctuate between 0 and 1.\footnote{When we say ``fluctuate" we have in mind ``in the real topology."
So we should remember that all our probabilistic models, e.g., Kolmogorov and von Mises, are rigidly
coupled to one very special topology, namely, the real one. Where is the origin of the real 
topological probability? In the real topology of  space-time? Do all our probabilistic models
describe just the reality of the real space-time? I discussed these problems at many occasions,
see, e.g., [31], [35]. In particular, I developed a $p$-adic probability model in which reality
($p$-adic reality) is associated with $p$-adic collectives-contexts, i.e., ``random sequences"
for which relative frequencies stabilize in the $p$-adic topology, [´31], [35].}

The absence of the statistical stabilization of relative frequencies for microparameters
does not at all contradict to the statistical stabilization of relative frequencies 
for macro-observables, see examples in [31].

In principle, chaotic fluctuations in the microworld may generate statistical stability on the 
macrolevel!

Neither it contradicts to realism. But we should distinguish INDIVIDUAL REALISM
(physical observables are objective properties of physical systems, i.e., mathematically
we can represent them as functions of HV, $a=a(\lambda))$ and KOLMOGOROV-BELL REALISM, namely, the existence
of the probability distribution.

Thus von Mises approach strongly differs  from Kolmogorov's approach. I disagree
with R. Gill who claimed that these approaches give the same consequences:
"Regarding to Kolmogorov and von Mises... I do not see any opposition between
alternative views of probability here," [16].

{\bf 2.3. Contextual probability.} The frequency approach is really a contextual probabilistic approach. Here we
work not with just one fixed collective-context, but with a few collective-contexts
by combining probabilities belonging to different collectives-contexts. Of course,
R. von Mises did not do so much in this direction. But in any case it was 
the important step compare with the Kolmogorov approach. Of course, Kolmogorov
also paid attention on contextuality of probabilities, see [19], [36]. But his ideology was:
for any fixed physical context we should choose a fixed Kolmogorov probability space 
and work in this space for ever! Even Kolmogorov-conditioning is just event-conditioning 
in one fixed Kolmogorov probability space -- so
for a fixed context. R. Gill and I. Helland paid my
attention to the fact that statisticians often consider families of Kolmogorov probability spaces
depending on some parameter $\sigma.$ Then by using statistics for random variables
they try to find this parameter. However, it is totally different ideology. I think that
the main role contextualism of probabilities play when we move from one context to another.
In really contextual probability theory we should be able to work not only in one fixed
context, but also to describe transformations of probabilities which are induced by context
transitions. Such a  contextualism can be called {\it Transition Contextualism} and traditional Kolmogorov
(and statistical) contextualism -- {\it Stationary Contextualism.}\footnote{There are some probabilistic
approaches which are based on conditional probabilities, see, e.g., Renyi [37] or Cox [38]. However, 
those approaches are still rigidly coupled to Stationary Contextualism. Therefore I disagree with L. Ballentine [1]
who claimed that quantum probability can be reduced to such a conditional probability --
he used the Cox-model. On the other hand, I appreciate Ballentine's investigations on conditional probabilistic
viewpoint of quantum probability. It was the first step -- Stationary Contextualism -- in the direction of quantum 
contextualism. I also remark that even earlier L. Accardi provided a detailed analysis of the role of Bayes
formula for conditional probability in understanding of quantum probability [39].} 

I think that one of the main distinguishing features of the quantum 
probabilistic formalism is the possibility to find dependence of probabilities on contexts. 
By changing a representation (the orthonormal basis) in a Hilbert space we change context. And 
the quantum formalism gives us the transformation of probabilities induced by such a context change.
Thus quantum probabilistic formalism is a transition-contextual formalism.

At the moment we do not have an intuitively clear transition-contextual probability theory.
Therefore I tried to proceed in the von Mises approach. I do not claim that this is the final
contextual probabilistic theory. But even by using the von Mises model I could do a lot to
clarify the probabilistic structure of quantum theory. As we have already seen, 
there are at least three contextual arguments against conclusions which J. Bell did on the basis
of his Kolmogorov model for the EPR-Bohm experiment: no counterfactual data, no independence of
collectives (so the absence of the  Bell-Clauser-Horne locality condition), the absence of physical
argument supporting Kolmogorov-Bell realism. Moreover, in the frequency
model, instead of Bell's (or CHSH) inequality, we obtain some  modifications  of this inequality,
see [40]. In general these modified inequalities do not contradict to experiments. Moreover, recently I derived 
the quantum EPR-Bohm correlations in the frequency approach (without to appeal to the Hilbert
space formalism), see [41].\footnote{Of course, if we are discussing just the Kolmogorov-Bell realism, then all our 
frequency (contextual) arguments are meaningless. But it seems not to be  the case!
It seems that when we discuss realism we have in mind INDIVIDUAL REALISM. Such a
realism can be described very well by the frequency probability model.
I also remark that by using the frequency (contextual) model we can easily 
resolve the GHZ-paradox, see [42]. Of course, we assume that GHZ were interested
to discuss realism and not just the Kolmogorov-Bell realism.}

{\bf 2.4. Quantum$\equiv$contextual?} My contextual probabilistic investigations  demonstrated 
that many things which are traditionally assigned to the quantum domain can be reproduced 
by using a contextual probabilistic model. In particular, the interference of probabilities
which is typically considered as one of the fundamental  features of quantum systems
can be obtained in a contextual model, see [43] on the frequency derivation of the interference.

The following question is of the great interest for me: 

\medskip

{\it Can ``quantum" be totally reduced to ``contextual"? Or ``quantum" is something 
more special than ``contextual"?}

\medskip

To investigate this question in more detail I considered a formal contextual probability model [5].
I did not try to provide a mathematical description of a context (if it is possible
at all in the general case). In such a formal model contexts are just some labels which
are assigned to probabilities, $P=P_{{\cal S}}.$ This is most general framework based on the well acceptable
postulate: 

\medskip

{\it All probabilities depend on contexts -- complexes of experimental physical conditions.}

\medskip

We can generalize von Mises slogan by saying: first context then probability. 
In this framework we can reproduce, e.g., the interference of probabilities. However, we immediately
see that the formal contextual probability model describes essentially larger set of context
transitions than quantum theory. In particular, I found that, besides the trigonometric
(``quantum") interference, there can appear the hyperbolic interference. The latter transformation
of probabilities also has a linear representation, namely, in a module over a Clifford algebra [44].

\medskip

{\bf ``Theorem".} {\it Quantum domain is just a proper subset of contextual domain.}

\medskip

The fundamental question (at least for me) is :
Which additional elements should we add to ``contextual" to obtain really ``quantum"?

Recently in the process of our discussions  with I. Volovich we understood that space-time does not 
present in the conventional axiomatics of quantum mechanics, see, e.g., von Neumann [45].
On the basis of our discussion I. Volovich presented a new system of axioms for quantum mechanics,
see [46]. Roughly speaking this is von Neumann's axiomatics with the additional space-time axiom.
However, in the process of further discussions  we understood that space-time
is nothing than a special context -- {\it space-time context}, see our quant-preprint [47].
According to the general ideology of the contextual approach probabilities should also depend
on the space-time context.  I do not think that space-time as itself is the ``fundamental quantum element."
The same space-time we also use in classical mechanics.

\section{``Fundamental quantum element"}

The contextual probabilistic investigations demonstrated that a rather special behaviour of
quantum probabilities is, in fact, nothing than contextual behaviour. Thus the Hilbert 
space calculus of probabilities is not the ``fundamental quantum element".
The fundamental is the very special 
form of time-evolution of probability distributions, namely, {\it Schr\"odinger evolution.} 
I do not afraid to say that the real quantum mechanics was discovered by E. Shr\"odinger [48]
and not by W. Heisenberg [49]. In particular, I do not consider {\it noncommutativity} as a 
fundamental really quantum feature. The noncommutative representation of physical observables 
in conventional quantum theory is just a sign of contextuality. In principle, such a representation
can arise in various classical physical models. Such examples were presented in our paper with
S. Kozyrev [50]. I think that there can be found hundreds of noncommutative classical models.
In particular, Heisenberg's {\it uncertainty relations} are just contextual statistical relations
describing dependence of dispersions on physical contexts. From this point of view 
{\it quantum logic} is, in fact, not quantum, but a {\it contextual logic.}

The main quantum mystery is 

\medskip

{\bf MYSTERY of  SCHR\"ODINGER REPRESENTATION.}

\medskip

I would like to thank L. Ballentine, K. Valiev, S. Albeverio, E. Beltrametti, L. Accardi, E. Loubenets, R. Gill,
T. Hida, D. Greenberger, S. Gudder,  W. De Muynck, J. Summhammer, P. Lahti, J-A. Larsson, 
H. Atmanspacher, 
B. Coecke, S. Aerts, A. Peres, A. Holevo,   L. Polley, A. Zeilinger, C Fuchs,  L. Hardy,
A. Plotnitsky, A. Shimony, R. Jozsa, J. Bub, C. Caves, K. Gustafsson, 
H. Bernstein, 
for fruitful (and rather critical) discussions on foundations of quantum mechanics.

I would like to thank B. Hiley  for  discussions on the role of space in 
quantum formalism (in the relation to the Bohmian mechnaics),  S. Molotkov  for discussions on 
the role of space-time in quantum information,  and  I. Volovich for discussions on
the space-time axiom of quantum theory. 

\medskip

{\bf References}

1. L. E. Ballentine,  Interpretations of probability and quantum theory.
Proc. Conf. {\it Foundations of Probability and Physics,} ed. A. Khrennikov.
{\it Q. Prob. White Noise Anal.}, {\bf 13}, 71-84, WSP, Singapore (2001).

L. E. Ballentine,  Probability theory in quantum mechanics. {\it American
J. of Physics}, {\bf 54,} 883-888 (1986).

2. L. Hardy, Quantum theory from intuitively reasonable axioms. 
Proc. Conf. {\it Quantum Theory: Reconsideration
of Foundations,} ed. A. Khrennikov. 
Ser. Math. Modelling, {\bf 2}, 117-130, V\"axj\"o Univ. Press (2002).

3. R. Gill, G. Weihs, A. Zeilinger, M. Zukowski, {\it Comment on "Exclusion of time in the theorem 
of Bell" by K. Hess and W. Philipp.} quant-ph/0204169 (2002).

R. Gill, Accardi contra Bell: the impossible coupling. In ed. M. Moore, C. Leger, and S. Froda. 
{\it Mathematical Statistics and Applications: Festschrift for Cnstance van Eedan.}
Lecture Notes-Monorgaph series, Hayward, Ca. Inst. Math. Stat., (2003).

4. S. P. Gudder, An approach to quantum probability. Proc. Conf.
{\it Foundations of Probability and Physics,} ed. A. Khrennikov.
Quantum Prob. White Noise Anal., {\bf 13}, 147-160, WSP, Singapore (2001).

5.  A. Yu. Khrennikov, {\it Contextual viewpoint to quantum stochastics.} hep-th/0112076.

A. Yu. Khrennikov, On foundations of quantum theory. 
Proc. Conf. {\it Quantum Theory: Reconsideration
of Foundations,} ed. A. Khrennikov. 
Ser. Math. Modelling,, {\bf 2}, 163-196,V\"axj\"o Univ. Press (2002).

6. I. Pitowsky, Range theorems for quantum probability and entanglement.
Ibid, {\bf 2}, 299-308, V\"axj\"o Univ. Press (2002).

7. J. Summhammer, Conservation of statistical information and quantum laws. 
Ibid, {\bf 2}, 367-384, V\"axj\"o Univ. Press (2002).

A. Khrennikov and J. Summhammer, Dialogue on classical and quantum between 
mathematician and experimenter. Ibid, {\bf 2}, 365-394, 
V\"axj\"o Univ. Press (2002).

8. L. Accardi and M. Regoli,  Locality and Bell's inequality.
Proc. Conf.{\it Foundations of Probability and Physics,} ed. A. Khrennikov.
{\it Q. Prob. White Noise Anal.}, {\bf 13}, 1-28, WSP, Singapore (2001).

9. W. De Baere, W. Struyve, Subquantum nonreproducibility and the complete local description of physical reality.
Proc. Conf. {\it Quantum Theory: Reconsideration
of Foundations,} ed. A. Khrennikov. 
Ser. Math. Modelling, {\bf 2}, 59-74, V\"axj\"o Univ. Press (2002).

10. W. M. De Muynck, Interpretations of quantum mechanics, and interpretations of 
violation of Bell's inequality. 
Proc. Conf. {\it Foundations of Probability and Physics,} ed. A. Khrennikov.
{\it Q. Prob. White Noise Anal.}, {\bf 13}, 95-114, WSP, Singapore (2001).

11. K. Hess and W. Philipp, {\it Einstein-separability, time related hidden 
parameters for correlated spins, and the theorem of Bell.} quant-ph/0103028;
{\it Proc. Nat. Acad. Sci. USA,} {\bf 98}, 14224 (2001); 
{\it Proc. Nat. Acad. Sci. USA,} {\bf 98}, 14228 (2001); 
{\it Europhys. Lett.}, {\bf 57}, 775 (2002).

12. A. Khrennikov, {\it Einstein and Bell, von Mises and Kolmogorov: reality,
locality, frequency and probability.} quant-ph/0006016 (2000).

13. I. Volovich, Quantum cryptography in space and Bell's theorem.
Proc. Conf. {\it Foundations of Probability and Physics,} ed. A. Khrennikov.
{\it Q. Prob. White Noise Anal.}, {\bf 13}, 364-372, WSP, Singapore (2001).

14. C. Fuchs, Quantum mechanics as quantum information (and only a little more).
Proc. Conf. {\it Quantum Theory: Reconsideration
of Foundations,} ed. A. Khrennikov. 
Ser. Math. Modelling, {\bf 2}, 463-543, V\"axj\"o Univ. Press (2002).

15.  G. `t Hooft, {\it How does God play dice? (Pre)-determinism at the Planck scale.}
hep-th/0104219 (2001).

G. `t Hooft, {\it Quantum mechanics and determinism.} hep-th/0105105 (2001).

16. A. Yu. Khrennikov, Ja. I. Volovich, Interference effect for probability distributions
of determinsitic particles. Proc. Conf. {\it Quantum Theory: Reconsideration
of Foundations,} ed. A. Khrennikov. Ser. Math. Modelling, {\bf 2}, 455-462,
V\"axj\"o Univ. Press (2002). quant-ph/0111159

A. Yu. Khrennikov, Ya. I.  Volovich, Discrete time leads to
quantum--like interference of deterministic particles,
Proc. Conf. {\it Quantum Theory: Reconsideration
of Foundations,} ed. A. Khrennikov. Ser. Math. Modelling, {\bf 2}, 441-454,
V\"axj\"o Univ. Press (2002). quant-ph/0203009.

17. R. Gill, Time, finite statistics, and Bell's fifth position.
To be published in Proc. Conf. {\it Foundations of Probability and Physics-2,} ed. A. Khrennikov.
Ser. Math. Modelling, {\bf 5},  V\"axj\"o Univ. Press (2003).
http://www.math.uu.nl/people/gill/Preprints/vaxjo.pdf

18.  R. von Mises,  {\it The mathematical theory of probability and
statistics.} Academic, London (1964).

19. A. N. Kolmogoroff, {\it Grundbegriffe der Wahrscheinlichkeitsrechnung.}
Springer Verlag, Berlin (1933); reprinted:
{\it Foundations of the Probability Theory}. 
Chelsea Publ. Comp., New York (1956).

20. P. A. M.  Dirac, {\it The Principles of Quantum Mechanics}
Oxford Univ. Press, (1930).

21. J. S. Bell, {\it Speakable and unspeakable in quantum mechanics.}
Cambridge Univ. Press (1987).

J. F. Clauser ,  A. Shimony, {\it Rep. Progr.Phys.,} 
{\bf 41} 1881-1901 (1978).

A. Aspect,  J. Dalibard,  G. Roger, 
{\it Phys. Rev. Lett.}, {\bf 49}, 1804-1807 (1982).

D. Home,  F. Selleri, {\it Nuovo Cim. Rivista,} {\bf 14},
2--176 (1991).

22. D. Bohm, {\it Quantum theory.} Prentice-Hall, Englewood Cliffs (1951).

23. A. Einstein, B. Podolsky, N. Rosen,  Phys. Rev., {\bf 47}, 777--780
(1935).

24.  A. Yu. Khrennikov, I. Volovich, {\it Einstein, Podolsky and Rosen versus
Bohm and Bell.} quant-ph/0211078.

25. S. Molotkov, Quantum teleportantion of a single-photon wave packet.
{\it Physics Lett. A}, {\bf 245}, 339-344 (1998).

S. Molotkov, Relativistic quantum cryptography on ``stopped" fotons.
{\it Letters JETF,} {\bf 76}, 79-85 (2002).

S. Molotkov and S. Nazin, Relativistic constraints on the distinguishability of 
orthogonal quantum states. {\it Atoms, Spectra, Radiation}, {\bf 94}, 1080-1087 (2002).

26. I. Volovich, Towards quantum information in space and time.Proc. Conf. {\it Quantum Theory: Reconsideration
of Foundations,} ed. A. Khrennikov. 
Ser. Math. Modelling, {\bf 2}, 423-440, V\"axj\"o Univ. Press (2002).

27. R. Feynman and A. Hibbs, {\it Quantum Mechanics and Path Integrals.}
McGraw-Hill, New-York (1965).

28. S. P. Gudder, {\it J. Math Phys.,} {\bf 25}, 2397- 2401 (1984).

29. I. Pitowsky, {\it Phys. Rev. Lett}, {\bf 48}, N.10, 1299-1302 (1982);
{\it Phys. Rev. D}, {\bf 27}, N.10, 2316-2326 (1983).

30. E. Beltrametti and G. Cassinelli,  The logic of quantum mechanics.
Addison-Wesley, Reading, Mass (1981).

31. A. Yu. Khrennikov, {\it Interpretations of Probability},
VSP Int. Sc. Publ., Utrecht (1999).

A. Yu. Khrennikov, Frequency analysis of the EPR-Bell argumentation. {\it Foundations of Physics,}
{\bf 32,} 1159-1174 (2002).

A. Khrennikov, {\it Comment on Hess-Philipp anti-Bell and 
Gill-Weihs-Zeilinger-Zukowski anti-Hess-Philipp
arguments.} quant-ph/0205022.

32.  M. van Lambalgen , Von Mises' definition of random sequences reconsidered.
{\it J. Symbolic Logic,} {\bf 52,} N. 3, (1987).

33.  H. P. Stapp, {\it  Phys. Rev.,} D, {\bf 3}, 1303-1320 (1971).

34. P. H. Eberhard, {\it Nuovo Cimento}, B, {\bf 46}, 392-400 (1978).

35. A.Yu. Khrennikov, {\it $p$-adic valued distributions in 
mathematical physics.} Kluwer Academic, Dordrecht, (1994).

A. Yu. Khrennikov, {\it Non-Archimedean analysis: quantum
paradoxes, dynamical systems and biological models.}
Kluwer Academic, Dordrecht (1997).

A. Yu. Khrennikov, Nonconventional viewpoint to elements of physical reality
based on nonreal asymptotics of relative frequencies. Proc. Conf.
{\it Foundations of Probability and Physics,} ed. A. Khrennikov.
{\it Q. Prob. White Noise Anal.}, {\bf 13}, 201-218, WSP, Singapore (2001).

36. A. N. Kolmogorov, The Theory of Probability. In: A.D.Alexandrov,
A.N.Kol\-mo\-go\-rov, M.A.Lavrent'ev (Eds.) {\it Mathematics, Its Content, Methods,
and Meaning,} {\bf 2,} M.I.T. Press  (1965).

37. Renyi ,  On a new axiomatics of probability theory. {\it Acta Mat. Acad. Sc. Hung.},
{\bf 6}, 285-335 (1955).

38. R. Cox, {\it The algebra of probable inference.} J. Hopkins Univ. Press, 
Baltimore (1961).

39. L. Accardi, The probabilistic roots of the quantum mechanical
paradoxes. {\it The wave--particle dualism. A tribute to Louis de Broglie on his 90th
Birthday.} (Perugia, 1982). Ed. S. Diner, D. Fargue, G. Lochak and F.
Selleri. D. Reidel Publ. Company, Dordrecht, 297--330 (1984).

40. A. Yu. Khrennikov, A perturbation of CHSH inequality induced by fluctuations 
of ensemble distributions. {\it J. of Math. Physics}, {\bf 41}, N.9, 5934-5944 (2000). 

41. A. Yu. Khrennikov, Local realist 
(but contextual) derivation of the EPR-Bohm correlations. quant-ph/0211073. 

42. A. Yu. Khrennikov, Contextualist viewpoint to 
Greenberger-Horne-Zeilinger paradox. {\it Phys. Lett.}, A, {\bf 278}, 307-314 (2001).

43. A. Yu. Khrennikov, Linear representations of probabilistic transformations induced by
context transitions. {\it J. Phys.A: Math. Gen.,} {\bf 34}, 9965-9981 (2001).
quant-ph/0105059.

44. A. Yu. Khrennikov, {\it Hyperbolic quantum mechanics.} quant-ph/0101002 (2000).

45. J. von Neumann, {\it Mathematical foundations of quantum mechanics.}
Princeton Univ. Press (1955).

46. I. Volovich, {\it Seven principles of quantum mechanics.} quant-ph/0212126 (2001).

47. A. Yu. Khrennikov, I. Volovich, Local Realism, Contextualism  and
Loopholes in Bell`s Experiments. quant-ph/0212127.

48.  E. Schr\"odinger, {\it Ann. Physik,} {\bf 79}, 361 (1926).

 E. Schr\"odinger, {\it Ann. Physik,} {\bf 79}, 489 (1926).

49. W. Heisenberg, {\it Zeits. f\"ur Physik,} {\bf 33},  879 (1925).

50. A. Yu. Khrennikov and S. Kozyrev, Noncommutative probability in classical systems.
quant-ph/ (2001).

\end{document}